\documentclass[aps,prl,preprint,tightenlines,superscriptaddress,showpacs,byrevtex]{revtex4}

\usepackage{graphicx} % Include figure files
\usepackage{dcolumn}  % Align table columns on decimal point
\usepackage{epsfig}
\usepackage{amssymb}

\graphicspath{{figs/}}

\begin{document}

\preprint{\vbox{ \hbox{   }
                 \hbox{BELLE-CONF-0318}
                 \hbox{EPS Parallel Session: 3, 10, \& 12}
                 \hbox{EPS-ID 536}
                 \hbox{hep-ex/0307077}
}}

\title{ \quad\\[0.5cm] Studies of $B^0 \to \rho^\pm \pi^\mp$
  and Evidence of $B^0 \to \rho^0 \pi^0$}

%%% Paper:    (fill in)
%%% Journal:  Summer 2003 conference proceedings (Physical Review format)
%%% Contacts: (fill in)
%%% Each author is included unless he/she chooses to opt out of ALL papers.
%%% ====================================================================
%%% Click the RELOAD button on your web browser to see the updated file.
%%% ====================================================================
%%% Use \input{author} to insert this material into your latex file.
%%%%% Force institutions to appear in alphabetical order when typeset.
\affiliation{Aomori University, Aomori}
\affiliation{Budker Institute of Nuclear Physics, Novosibirsk}
\affiliation{Chiba University, Chiba}
\affiliation{Chuo University, Tokyo}
\affiliation{University of Cincinnati, Cincinnati, Ohio 45221}
\affiliation{University of Frankfurt, Frankfurt}
\affiliation{Gyeongsang National University, Chinju}
\affiliation{University of Hawaii, Honolulu, Hawaii 96822}
\affiliation{High Energy Accelerator Research Organization (KEK), Tsukuba}
\affiliation{Hiroshima Institute of Technology, Hiroshima}
\affiliation{Institute of High Energy Physics, Chinese Academy of Sciences, Beijing}
\affiliation{Institute of High Energy Physics, Vienna}
\affiliation{Institute for Theoretical and Experimental Physics, Moscow}
\affiliation{J. Stefan Institute, Ljubljana}
\affiliation{Kanagawa University, Yokohama}
\affiliation{Korea University, Seoul}
\affiliation{Kyoto University, Kyoto}
\affiliation{Kyungpook National University, Taegu}
\affiliation{Institut de Physique des Hautes \'Energies, Universit\'e de Lausanne, Lausanne}
\affiliation{University of Ljubljana, Ljubljana}
\affiliation{University of Maribor, Maribor}
\affiliation{University of Melbourne, Victoria}
\affiliation{Nagoya University, Nagoya}
\affiliation{Nara Women's University, Nara}
\affiliation{National Kaohsiung Normal University, Kaohsiung}
\affiliation{National Lien-Ho Institute of Technology, Miao Li}
\affiliation{Department of Physics, National Taiwan University, Taipei}
\affiliation{H. Niewodniczanski Institute of Nuclear Physics, Krakow}
\affiliation{Nihon Dental College, Niigata}
\affiliation{Niigata University, Niigata}
\affiliation{Osaka City University, Osaka}
\affiliation{Osaka University, Osaka}
\affiliation{Panjab University, Chandigarh}
\affiliation{Peking University, Beijing}
\affiliation{Princeton University, Princeton, New Jersey 08545}
\affiliation{RIKEN BNL Research Center, Upton, New York 11973}
\affiliation{Saga University, Saga}
\affiliation{University of Science and Technology of China, Hefei}
\affiliation{Seoul National University, Seoul}
\affiliation{Sungkyunkwan University, Suwon}
\affiliation{University of Sydney, Sydney NSW}
\affiliation{Tata Institute of Fundamental Research, Bombay}
\affiliation{Toho University, Funabashi}
\affiliation{Tohoku Gakuin University, Tagajo}
\affiliation{Tohoku University, Sendai}
\affiliation{Department of Physics, University of Tokyo, Tokyo}
\affiliation{Tokyo Institute of Technology, Tokyo}
\affiliation{Tokyo Metropolitan University, Tokyo}
\affiliation{Tokyo University of Agriculture and Technology, Tokyo}
\affiliation{Toyama National College of Maritime Technology, Toyama}
\affiliation{University of Tsukuba, Tsukuba}
\affiliation{Utkal University, Bhubaneswer}
\affiliation{Virginia Polytechnic Institute and State University, Blacksburg, Virginia 24061}
\affiliation{Yokkaichi University, Yokkaichi}
\affiliation{Yonsei University, Seoul}
  \author{K.~Abe}\affiliation{High Energy Accelerator Research Organization (KEK), Tsukuba} % KEK
  \author{K.~Abe}\affiliation{Tohoku Gakuin University, Tagajo} % TohokuGakuin
  \author{N.~Abe}\affiliation{Tokyo Institute of Technology, Tokyo} % TIT
  \author{R.~Abe}\affiliation{Niigata University, Niigata} % Niigata
  \author{T.~Abe}\affiliation{High Energy Accelerator Research Organization (KEK), Tsukuba} % KEK
  \author{I.~Adachi}\affiliation{High Energy Accelerator Research Organization (KEK), Tsukuba} % KEK
  \author{Byoung~Sup~Ahn}\affiliation{Korea University, Seoul} % Korea
  \author{H.~Aihara}\affiliation{Department of Physics, University of Tokyo, Tokyo} % Tokyo
  \author{M.~Akatsu}\affiliation{Nagoya University, Nagoya} % Nagoya
  \author{M.~Asai}\affiliation{Hiroshima Institute of Technology, Hiroshima} % Hiroshima
% ...
% Lausanne
\collaboration{The Belle Collaboration}

\noaffiliation

\begin{abstract}
  We report studies of $B^0 \to \pi^+ \pi^- \pi^0$, 
  using $78\ {\rm fb}^{-1}$ of data collected at 
  the $\Upsilon(4S)$ resonance with the Belle detector 
  at the KEKB asymmetric $e^+e^-$ storage ring. 
  We measure the branching fraction for $B^0 \to \rho^\pm \pi^\mp$ to be
  ${\mathcal B}\left( B^0 \to \rho^\pm \pi^\mp \right) = 
  \left( 29.1 ^{+5.0}_{-4.9}({\rm stat}) \pm 4.0({\rm syst}) \right) 
  \times 10^{-6}$, and find an untagged charge asymmetry
  ${\mathcal A} =
  -0.38^{+0.19}_{-0.21} ({\rm stat}) ^{+0.04}_{-0.05} ({\rm syst})$.
  We find the first evidence for $B^0 \to \rho^0 \pi^0$ with 
  $3.1\sigma$ statistical significance and with a branching fraction of
  ${\mathcal B}\left( B^0 \to \rho^0\pi^0 \right) =
  \left( 6.0 ^{+2.9}_{-2.3} ({\rm stat}) \pm 1.2 ({\rm syst}) \right)
  \times 10^{-6}$.
\end{abstract}

\maketitle

\tighten

{\renewcommand{\thefootnote}{\fnsymbol{footnote}}}
\setcounter{footnote}{0}

There is a large amount of interest in $B^0 \to \pi^+\pi^-\pi^0$ 
decays~\cite{cc}.
The three body final state is expected to be dominated 
by the quasi-two body decays $B^0 \to \rho^{\pm} \pi^{\mp}$, 
the branching fraction of which has been previously measured to be 
${\mathcal B}\left( B^0 \to \rho^{\pm} \pi^{\mp} \right) = 
\left( 20.8 ^{+6.0+2.8}_{-6.3-3.1} \right) \times 10^{-6}$ 
by Belle~\cite{ascelin}, 
from a data sample of $29.4~{\rm fb}^{-1}$,
corresponding to $31.9 \times 10^6$ $B\bar{B}$ pairs.
Recently BaBar have announced a preliminary measurement of 
${\mathcal B}\left( B^0 \to \rho^{\pm} \pi^{\mp} \right) = 
\left( 22.6 \pm 1.8 \pm 2.2 \right) \times 10^{-6}$~\cite{babar_rhopi},
from a data sample corresponding to $89 \times 10^6$ $B\bar{B}$ pairs.
Additionally, they have performed a time-dependent analysis on 
the quasi-two body signal candidates.
Such an analysis can 
obtain information about the Unitary Triangle angle $\phi_2$~\cite{phi2},
which may be useful to help interpret the results of 
time-dependent $B^0 \to \pi^+\pi^-$ analyses~\cite{pipi}.

In addition to $\rho^{\pm} \pi^{\mp}$,
the $\pi^+\pi^-\pi^0$ final state can be accessed via 
$B^0 \to \rho^0 \pi^0$ decay.  
Analogous to $B^0 \to \pi^0\pi^0$ amongst the $\pi\pi$ final states,
${\mathcal B} \left( B^0 \to \rho^0 \pi^0 \right)$
can be used to limit the possible contribution from penguin diagrams.
Note that the penguin pollution in $\rho^{\pm} \pi^{\mp}$ 
is expected to be smaller than that in $\pi^+\pi^-$;
a hypothesis which is supported by comparing the ratio
${\mathcal B} \left( B^0 \to \rho^{\pm}K^{\mp} \right) /
{\mathcal B} \left( B^0 \to \rho^{\pm}\pi^{\mp} \right)$
to
${\mathcal B} \left( B^0 \to \pi^{\pm}K^{\mp} \right) /
{\mathcal B} \left( B^0 \to \pi^{\pm}\pi^{\mp} \right)$.
Currently the best upper limit is
${\mathcal B}\left( B^0 \to \rho^0 \pi^0 \right) < 
5.3 \times 10^{-6}$~\cite{ascelin}, 
whilst most theoretical estimates of this branching fraction are
${\mathcal O}\left(10^{-6}\right)$ or lower.

An unambiguous measurement of $\phi_2$ can, in principle, be made 
from the Dalitz plot of $B^0 \to \pi^+\pi^-\pi^0$~\cite{snyder_quinn}.
It is also possible to observe direct $CP$ violation from a population
asymmetry in the untagged Dalitz plot distribution~\cite{gardner}.
Furthermore, other resonant contributions 
to the $\pi^+\pi^-\pi^0$ final state are possible;
recently there has been particular theoretical interest in
$B^0 \to \sigma \pi^0$~\cite{sigma}.

With very large statistics, and sophisticated analysis techniques,
it would be possible to address all these open questions regarding 
$B^0 \to \pi^+\pi^-\pi^0$, using a time-dependent Dalitz plot analysis.
While we cannot as yet achieve this ambitious goal,
the results presented here represent milestones towards this objective.
We present an updated measurement of the $B^0 \to \rho^{\pm}\pi^{\mp}$
branching fraction, using event selection that can provide a sample
for a quasi-two body time-dependent analysis. 
We also present the first evidence for $B^0 \to \rho^0 \pi^0$.

The analysis is based on a 78~$\rm fb^{-1}$ data sample 
containing $85 \times 10^6$ $B$ meson pairs collected
with the Belle detector at
the KEKB asymmetric-energy $e^+e^-$ collider~\cite{KEKB}. 
KEKB operates at the $\Upsilon(4S)$ resonance 
($\sqrt{s}=10.58~{\rm GeV}$) with
a peak luminosity that exceeds
$1\times 10^{34}~{\rm cm}^{-2}{\rm s}^{-1}$.

The Belle detector is a large-solid-angle magnetic spectrometer that
consists of a three-layer silicon vertex detector (SVD),
a 50-layer central drift chamber (CDC), 
an array of aerogel threshold \v{C}erenkov counters (ACC), 
a barrel-like arrangement of time-of-flight scintillation counters (TOF), 
and an electromagnetic calorimeter comprised of CsI(Tl) crystals (ECL) 
located inside a super-conducting solenoid coil 
that provides a 1.5~T magnetic field.  
An iron flux-return located outside of the coil is instrumented 
to detect $K_L$ mesons and to identify muons (KLM).  
The detector is described in detail elsewhere~\cite{Belle}.

Charged tracks are required to originate from the interaction point 
and have transverse momenta greater than $100~{\rm MeV}/c$. 
To identify tracks as charged pions,
we combine $dE/dx$ information from the CDC,
pulse height information from the ACC and 
timing information from the TOF into
pion/kaon likelihood variables ${\mathcal L}(\pi/K)$.
We then require
${\mathcal L}(\pi)/\left( {\mathcal L}(\pi) + {\mathcal L}(K)\right) > 0.6$.
Additionally, tracks which are consistent with an electron hypothesis
are rejected.

Neutral pion candidates are reconstructed from photon pairs.
Photon candidates are selected with a mininum energy requirement 
of $50~{\rm MeV}$ in the barrel region of the ECL,
and $100~{\rm MeV}$ in its endcap.
The $\pi^0$ candidates are required to have momenta greater than 
$200~{\rm MeV}$ in the lab frame,
and a $\gamma\gamma$ invariant mass in the range
$0.118 < M_{\gamma\gamma}/{\rm GeV}/c^2 < 0.150$.
Additionally, we require
$\left| \cos(\theta^{\pi^0}_{\rm hel}) \right| < 0.95$,
where $\theta^{\pi^0}_{\rm hel}$ is defined as 
the angle between one photon's flight direction in the $\pi^0$ rest frame 
and the flight direction of $\pi^0$ with respect to the lab frame,
and make a loose requirement on the $\chi^2$ of a 
$\pi^0$ mass-constrained fit of $\gamma\gamma$.

$B$ candidates are selected using two kinematic variables:
the beam-constrained mass
$M_{bc}\equiv \sqrt{E^2_{\rm beam}-P^2_B}$ 
and the energy difference $\Delta E \equiv E_B - E_{\rm beam}$.
Here, $E_B$ and $P_B$ are the reconstructed energy and momentum 
of the $B$ candidate in the center of mass (CM) frame, 
and $E_{\rm beam}$ is the average beam energy in the CM frame. 
$B$ candidates with $M_{bc} > 5.2~{\rm GeV}/c^2$ and
$-0.30 < \Delta E / {\rm GeV} < 0.20$ are selected.
We further define the signal regions: 
$M_{bc} > 5.27~{\rm GeV}/c^2$ and
$-0.10 < \Delta E / {\rm GeV} < 0.08$.

To select $\rho^\pm\pi^\mp$ from 3-body $\pi^+\pi^-\pi^0$, 
we select candidates with an invariant $\pi \pi^0$ mass in the range
$\left| M_{\pi \pi^0} - M_{\rho} \right| < 0.20~{\rm GeV}/c^2$,
and $\rho$ helicity $\theta^{\rho}_{\rm hel}$,
 defined as the angle between the charged pion direction in the 
$\rho$ rest frame and the $\rho$ direction in the 
$B$ rest frame~\cite{helicity},
in the range 
$\left| \cos \theta^{\rho}_{\rm hel} \right| > 0.5$.

The dominant background to $B^0 \to \pi^+\pi^-\pi^0$
comes from continuum events, $e^+e^-\to q\bar{q}$ ($q = u, d, s, c$).
Since these tend to be jet-like, 
whilst $B\bar{B}$ events tend to be spherical,
we use event shape variables to discriminate between the two.
We combine five modified Fox-Wolfram moments~\cite{fox-wolfram}
into a Fisher discriminant;
the coefficients are then tuned to maximize the separation between
signal and continuum events.
We further define $\theta_B$ as the angle of the reconstructed $B$
candidate with respect to the beam direction.
Signal events have a distribution proportional to $\sin^2(\theta_B)$,
whilst continuum events are flatly distributed in $\cos(\theta_B)$.
We combine the output of the Fisher discriminant 
with $\cos(\theta_B)$ into signal/background likelihood variables, 
${\mathcal L}_{s/b}$.
We find the optimum selection requirement by maximizing
$S/\sqrt{S+B}$, where $S$ and $B$ are respectively the expected numbers of
signal and background events in the signal region.
We use our measured branching fraction of
$B^0 \to \rho^\pm\pi^\mp$~\cite{ascelin} as input,
and find the optimum requirement is
${\mathcal L}_s/\left( {\mathcal L}_s + {\mathcal L}_b \right) > 0.8$.

If more than one candidate remains in any event, 
that with the smallest $\chi^2_{\rm vtx} + \chi^2_{\pi^0}$ is selected,
where $\chi^2_{\rm vtx}$ is 
the $\chi^2$ of a vertex-constrained fit of $\pi^+\pi^-$,
and $\chi^2_{\pi^0}$ is 
that from a $\pi^0$ mass-constrained fit of $\gamma\gamma$.

We obtain the signal yield using a binned fit to the $\Delta E$ distribution,
and cross-check the result by fitting the $M_{\rm bc}$ distribution.
When fitting one variable, 
candidates are required to be in the signal region of the other.
The signal probability density functions (PDFs) 
are obtained from Monte Carlo (MC); 
a Crystal Ball~\cite{crystalball} lineshape 
plus a Gaussian is used for $\Delta E$, 
whilst the $M_{\rm bc}$ distribution is described by a Gaussian.
The Gaussian in the $\Delta E$ PDF accounts for poorly reconstructed 
low momentum neutral pions.
The $\Delta E$ width is calibrated using an inclusive 
$D^*$ sample ($D^{*+} \to D^0 \pi^+$, $D^0 \to K^-\pi^+\pi^0$)
whilst the $\Delta E$ and $M_{bc}$ peak positions are adjusted 
according to a data sample of $B^+\to D^0 \pi^+$ with $D^0\to K^-\pi^+\pi^0$.

The dominant background is from continuum events.
The $\Delta E$ distribution for these events is described by
a Chebyshev polynomial, 
whilst the $M_{\rm bc}$ shape is given by the ARGUS function~\cite{argus}.
The parameters of these functions are determined from fitting a large
continuum MC sample.

Background is also possible from generic $b \to c$ transitions;
in this case the shape is hard to describe by a functional form and
so we use a smoothed histogram.
The $\Delta E$ distribution of background from the charmless decay
$B^+ \to \rho^+ \rho^0$ has a similar shape to the generic $b \to c$,
so we combine these components,
with the relative normalization fixed according to our recent measurement
of ${\mathcal B}\left( B^+ \to \rho^+ \rho^0 \right)$~\cite{jingzhi},
and allow the overall normalization to float in the $\Delta E$ fit.
In the $M_{\rm bc}$ fit, 
these backgrounds cannot be distinguished from signal,
and the normalization is fixed.

There are other possible backgrounds from charmless $B$ decays,
with distinctive $\Delta E$ shapes.
$B^0 \to \rho^{\pm}K^{\mp}$ has a similar shape to the signal, 
but a shifted peak due to the misidentification of the kaon as a pion.
The normalization of this component is fixed according to
our recent measurement~\cite{paoti}.
Contributions from $\pi^+ \rho^0$, $hh$ and $h\pi^0$ ($h = \pi^{\pm},K^{\pm}$)
final states are scaled according to the most recent measurements 
of their branching fractions~\cite{ascelin,hh,pdg}, 
then combined into a smoothed histogram.
The normalization of this component is then fixed in the fit.

\begin{figure}
  \hbox to \hsize{
    \hss
    \includegraphics[width=0.45\hsize]{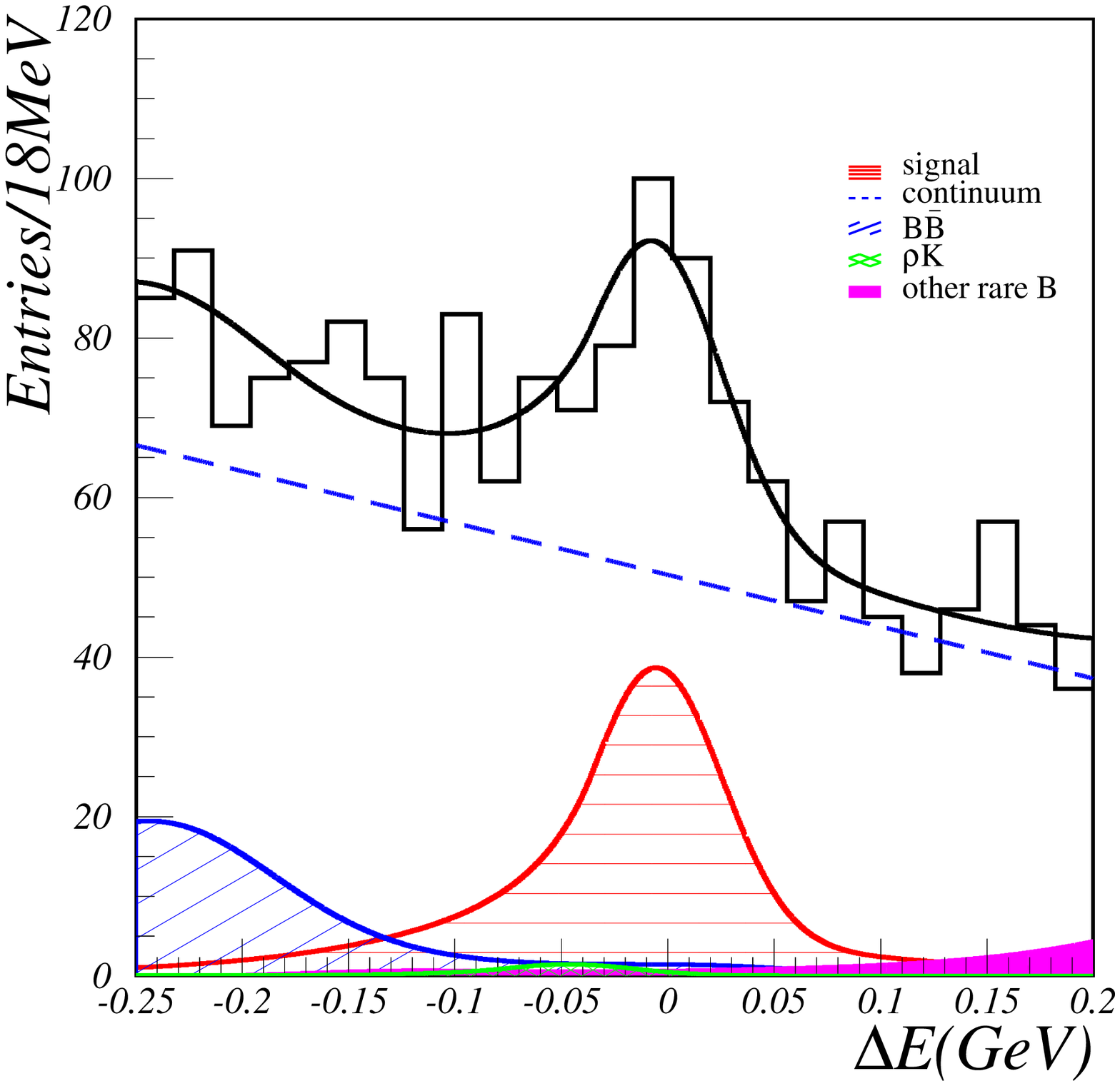}
    \hss
    \includegraphics[width=0.45\hsize]{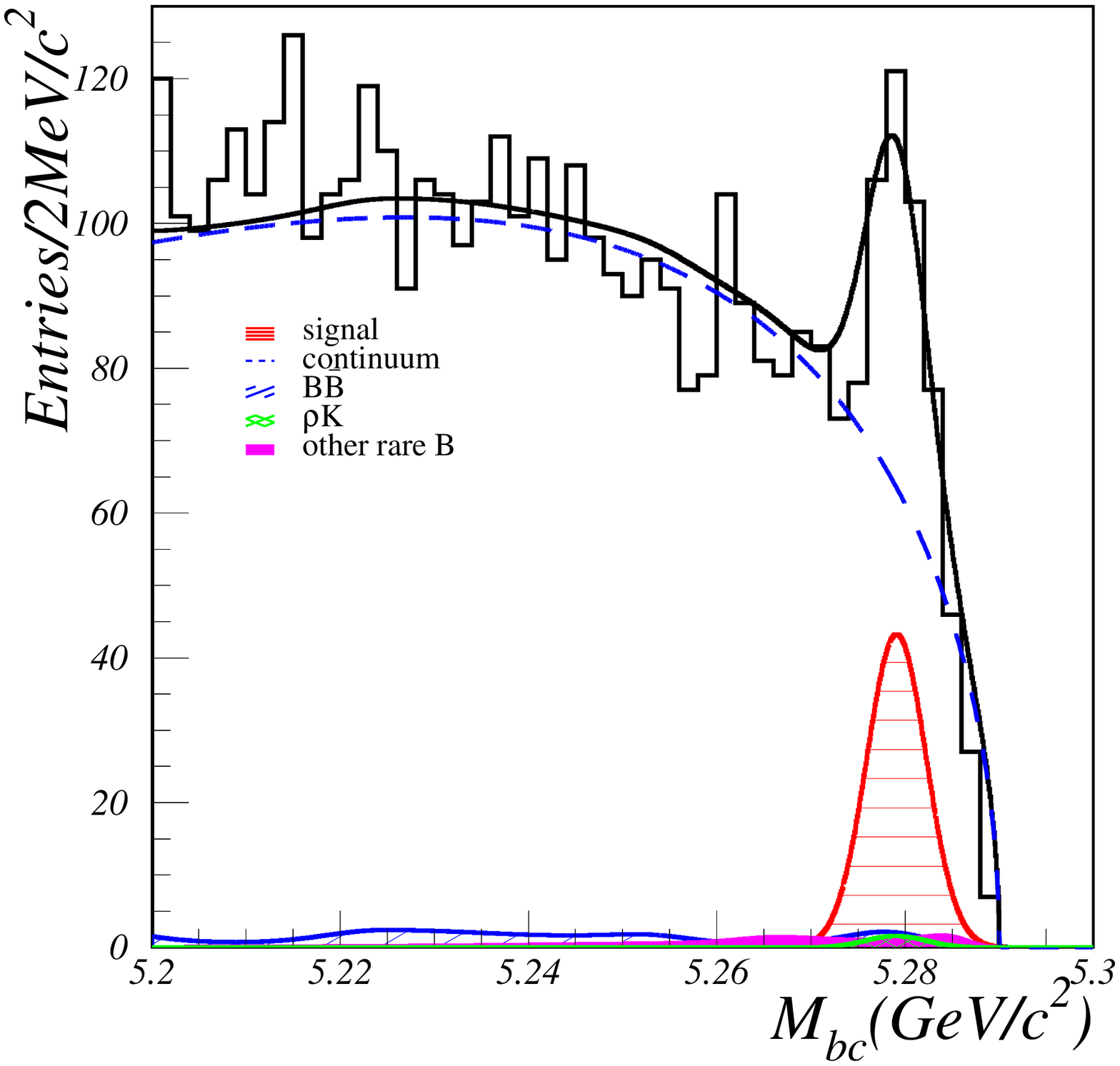}
    \hss
  }
  \caption{
    Results of fits to (left) $\Delta E$ and (right) $M_{\rm bc}$.
    Contributions from signal (horizontally hatched),
    continuum (dashed line) and $b \to c + \rho\rho$ 
    (diagonally hatched; denoted as $B\bar{B}$ in the legend)
    can be seen.
    The contributions from $\rho K$ and other rare decays are small
    and hard to make out.
    The sums of contributions are shown as solid lines.
  }
  \label{fig:rhopmpimp_fit}
\end{figure}

The results of the fits to $\Delta E$ and $M_{\rm bc}$ are shown in
Fig.~\ref{fig:rhopmpimp_fit}.
From the $\Delta E$ fit we find a yield of $257.9^{+44.0}_{-43.2}$
signal events, with a significance of $6.3\sigma$.

The significance is defined as 
$\sqrt{-2\ln({\mathcal L}_0/{\mathcal L}_{\rm max})}$, 
where ${\mathcal L}_{\rm max}$ (${\mathcal L}_0$)
denotes the likelihood with the signal yield at its nominal value 
(fixed to zero).

From the $M_{\rm bc}$ fit we find a signal yield of $177.7^{+24.7}_{-24.0}$.
Taking into account the different efficiencies of the 
$\Delta E$/$M_{\rm bc}$ signal region selections, 
this result is consistent with the $\Delta E$ yield.

\begin{figure}
  \hbox to \hsize{
    \hss
    \includegraphics[width=0.45\hsize]{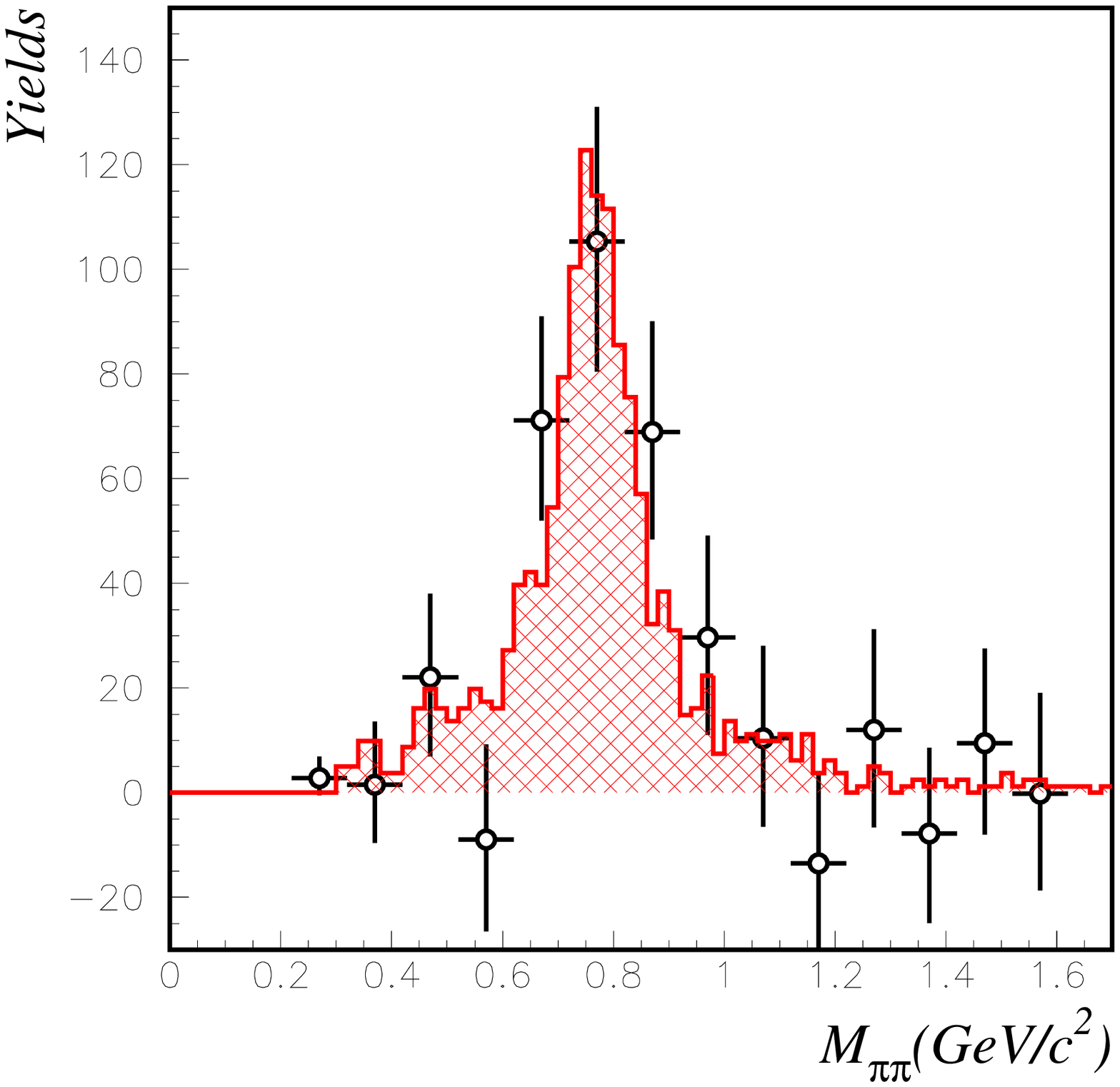}
    \hss
    \includegraphics[width=0.45\hsize]{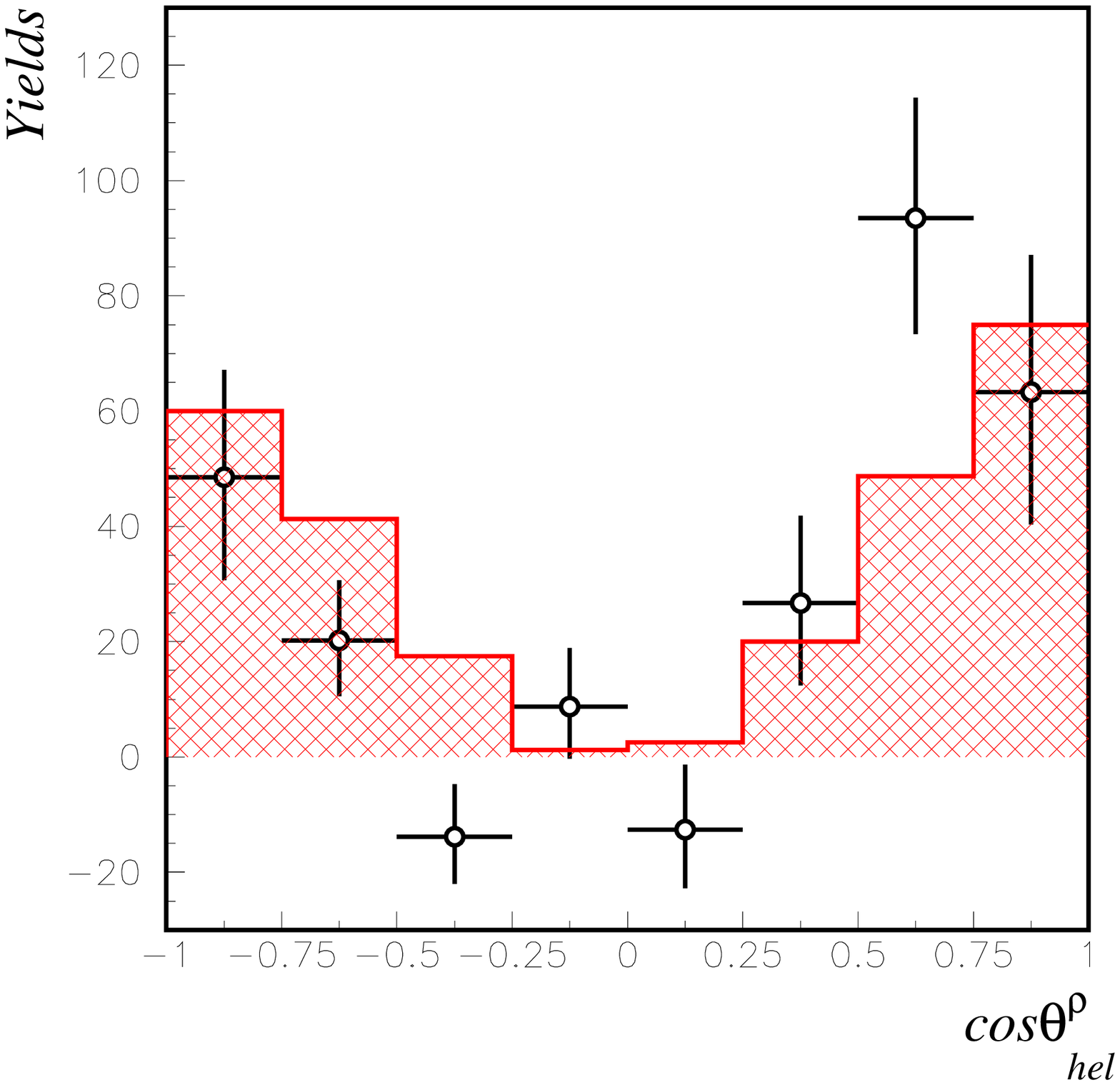}
    \hss
  }
  \caption{
    Yield of signal events from $\Delta E$ fits in bins of
    (left) $M_{\pi \pi^0}$ (right) $\cos \theta^{\rho}_{\rm hel}$.
    The shaded histogram shows the shapes expected from
    $B^0 \to \rho^{\pm}\pi^{\mp}$ Monte Carlo.
  }
  \label{fig:rhopmpimp_check}
\end{figure}

To check the events in the signal peak are 
$B^0\to \rho^\pm\pi^\mp$ events, 
and not from some other contribution to the $\pi^+\pi^-\pi^0$ final state,
we relax the $M_{\pi \pi^0}$ and $\cos \theta^{\rho}_{\rm hel}$
criteria in turn,
and perform the $\Delta E$ fit in bins of $M_{\pi \pi^0}$,
and in bins of $\cos \theta^{\rho}_{\rm hel}$.
The resulting distributions are shown in Fig.~\ref{fig:rhopmpimp_check}.
Clearly, the signal is consistent with being entirely due
to $B^0 \to \rho^{\pm}\pi^{\mp}$ decays.

To extract the branching fraction,
we measure the reconstruction efficiency from MC and correct for 
a discrepancy between data and MC in the pion identification requirement.
This correction is obtained from an inclusive $D^*$ control sample
($D^{*+} \to D^0 \pi^+$, $D^0 \to K^- \pi^+$),
and is applied in bins of track momentum and polar angle.
After this correction, the reconstruction efficiency is $10.4\%$.

We calculate systematic errors from the following sources:
PDF shapes $^{+11.7}_{-12.0}\%$ (by varying parameters by $\pm 1 \sigma$);
continuum rejection $\pm 6.5\%$
(by comparing the efficiency of the selection
between data and MC for the $B^+ \to D^0\pi^+$ control sample);
$\pi^0$ reconstruction efficiency $\pm 4.8\%$
(by comparing the yields of 
$\eta \to \pi^0\pi^0\pi^0$ and $\eta\to \gamma \gamma$
between data and MC);
tracking finding efficiency $\pm 2.0\%$
(from a study of partially reconstructed $D^*$ decays);
pion identification efficiency $\pm 0.8\%$
(using the method described above).
The contributions are summed in quadrature 
to obtain a total systematic error of $^{+13.6}_{-13.8}\%$.

We measure the branching fraction for $B^0 \to \rho^\pm \pi^\mp$ to be 
\begin{equation}
  \nonumber
  {\mathcal B}\left( B^0 \to \rho^\pm \pi^\mp \right) = 
  \left( 29.1 ^{+5.0}_{-4.9}({\rm stat}) \pm 4.0({\rm syst}) \right) 
  \times 10^{-6}.
\end{equation}

\begin{figure}
  \hbox to \hsize{
    \hss
    \includegraphics[width=0.45\hsize]{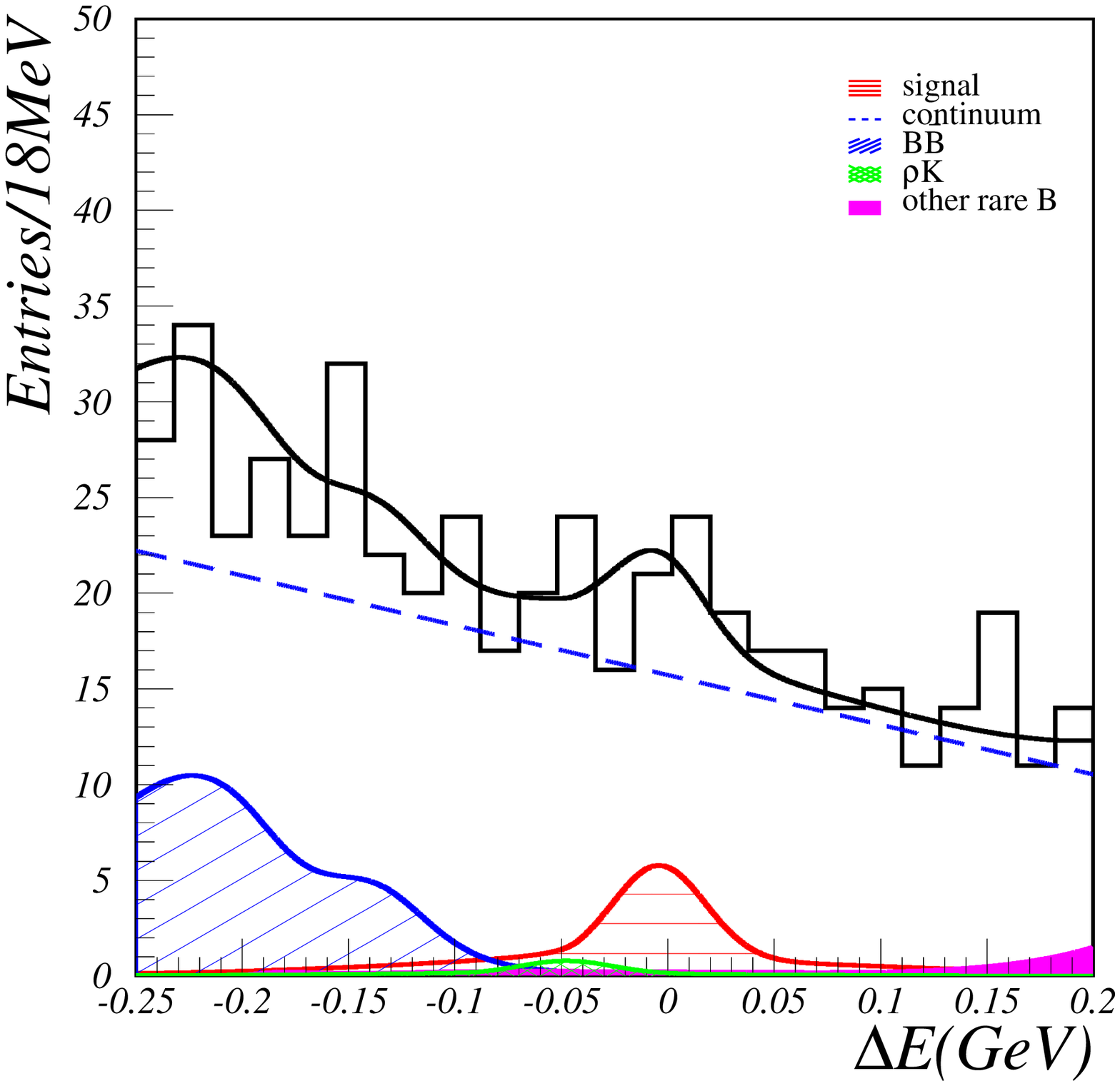}
    \hss
    \includegraphics[width=0.45\hsize]{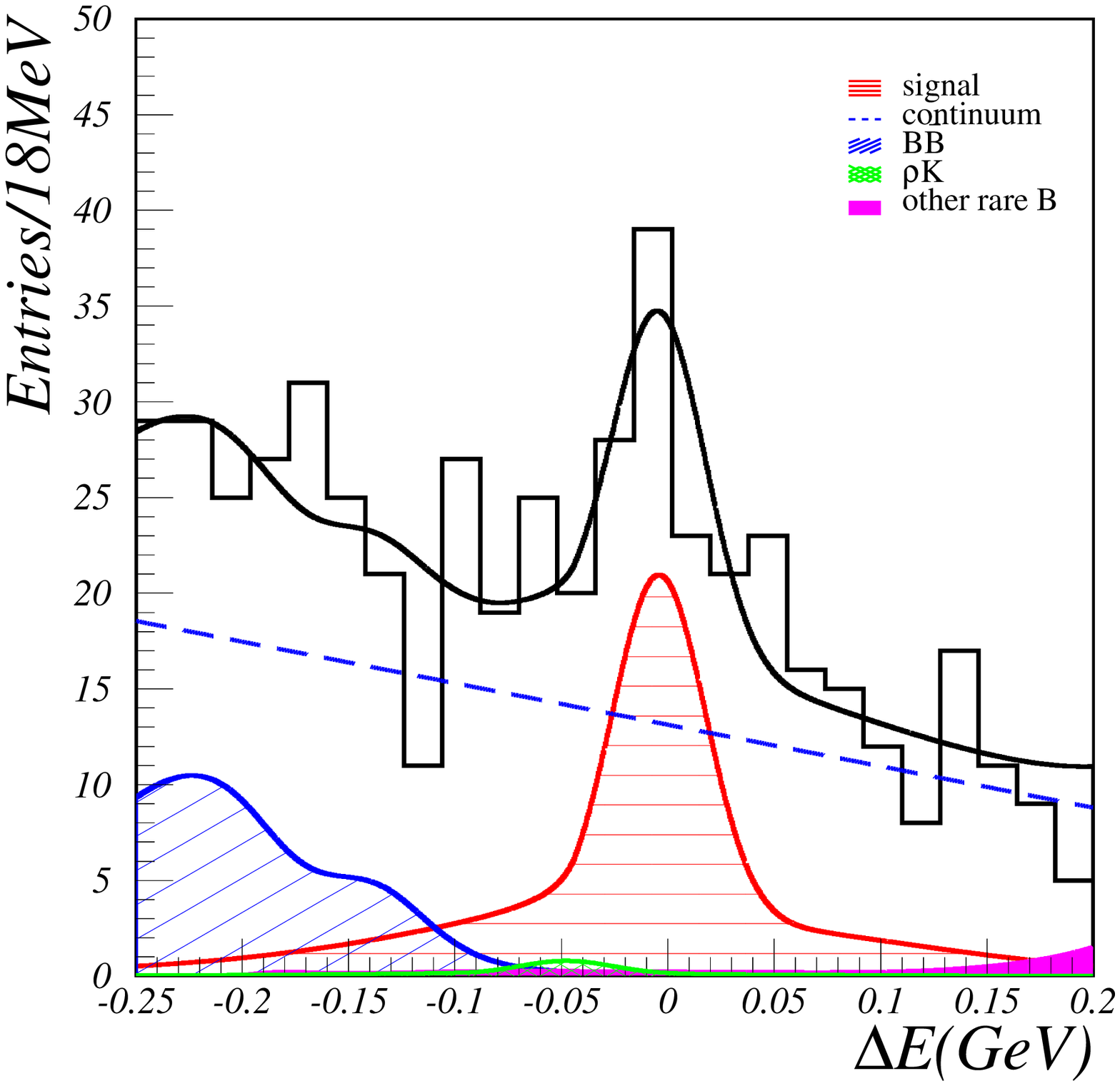}
    \hss
  }
  \caption{
    $\Delta E$ fit results for 
    (left) $B\to \rho^+ \pi^-$ and (right) $B\to \rho^- \pi^+$.
    The components have the same meaning as Fig.~\ref{fig:rhopmpimp_fit}.
  }
  \label{fig:rhopmpimp_asym}
\end{figure}

A difference in the untagged decay rates to $\rho^+\pi^-$ and $\rho^-\pi^+$
would indicate direct $CP$ violation~\cite{gardner}.
In order to measure this untagged asymmetry,
we remove candidate events with ambiguous $\rho$ charge assignment,
{\it i.e.} events in the regions of the Dalitz plot where
more than one of the combinations $\pi^+ \pi^0$, $\pi^-\pi^0$, $\pi^+\pi^-$
are consistent with having originated from a $\rho$ resonance.
These are the regions where interference effects are 
strongest~\cite{snyder_quinn}.
We fit the $\Delta E$ distributions for 
$B^0 \to \rho^+ \pi^- + \bar{B}^0 \to \rho^+ \pi^-$ 
(denoted as $B \to \rho^+ \pi^-$) and 
$B^0 \to \rho^- \pi^+ + \bar{B}^0 \to \rho^- \pi^+$ ($B \to \rho^- \pi^+$)
candidates separately.
The results are shown in Fig.~\ref{fig:rhopmpimp_asym}.
We find $36.7^{+15.3}_{-14.3}$ $B \to \rho^+ \pi^-$ events,
and $81.5^{+16.8}_{-16.0}$ $B \to \rho^- \pi^+$ events.
The charge asymmetry is calculated as
\begin{equation}\nonumber
  {\mathcal A} =
  \frac
  { N(B\to \rho^+\pi^-) - N(B\to \rho^-\pi^+) }
  { N(B\to \rho^+\pi^-) + N(B\to \rho^-\pi^+) }
  =  -0.38^{+0.19}_{-0.21} ({\rm stat}) ^{+0.04}_{-0.05} ({\rm syst}),
\end{equation}
where the systematic error is estimated by varying the PDFs used in the fit,
allowing for asymmetry in the shape and normalization of the continuum
component, and allowing for charge dependence in the efficiency.

\vspace{2ex}

In order to obtain good sensitivity to the as yet unobserved decay
$B^0 \to \rho^0 \pi^0$, additional discrimination against the 
continuum background is required.
Therefore, a slightly different analysis procedure is followed.
The majority of selection requirements are similar to those described above,
however, neutral pion candidates are selected with a wider invariant mass
window of $0.100 < M_{\gamma\gamma}/{\rm GeV}/c^2 < 0.165$
to allow for the resolution of high momentum $\pi^0$s.
One additional requirement is that possible contributions to the
$\pi^+\pi^-\pi^0$ final state from $b \to c$ decays are explicitly vetoed.
Another is that in order to reduce the event multiplicity before 
the best candidate selection, a smaller window in 
$\left( \Delta E, M_{\rm bc} \right)$ is allowed.
The selection region is $-0.2 < \Delta E/{\rm GeV} < 0.4$, 
$M_{\rm bc} > 5.23~{\rm GeV}/c^2$.
Whilst these changes do not significantly affect the $\rho^0 \pi^0$ 
final state, they will be important in retaining a clean and unbiased 
$\pi^+\pi^-\pi^0$ Dalitz plot in the future.

In order to select $\rho^0\pi^0$ from the three-body 
$\pi^+\pi^-\pi^0$ candidates, we require
$0.50 < M_{\pi^+\pi^-}/{\rm GeV}/c^2 < 1.10$ and
$\left| \cos \theta^{\rho}_{\rm hel} \right| > 0.5$.
Contributions from $B^0 \to \rho^{\pm}\pi^{\mp}$ 
are explicitly vetoed by rejecting candidates 
which fall into the invariant mass window of 
$0.50 < M_{\pi^{\pm}\pi^0}/{\rm GeV}/c^2 < 1.10$.
This requirement also vetoes the region of the Dalitz plot
where the interference between $\rho$ resonances is strongest.

The most notable difference from the selection requirements described above
for the $\rho^{\pm}\pi^{\mp}$ final state is in the procedure to reject
continuum background.
Here, we make use of the additional discriminatory power provided by
flavour tagging.
In our published time-dependent analyses~\cite{cpv},
we define the variables $q$ and $r$ as being, respectively,
the more likely flavour of the other $B$ in the event
($B^0\ (q = +1)$ or $\bar{B}^0\ (q = -1)$),
and a measure of the confidence that the $q$ prediction is correct.
As a corollary, events with a high value of $\left|qr\right|$
are well-tagged as either $B^0$ or $\bar{B}^0$,
and hence are unlikely to originate from continuum processes.
Moreover, since the flavour tagging algorithm relies
on particle identification information, 
it is unlikely that there is any strong correlation with 
any of the topological variables used above to separate signal from continuum.

\begin{figure}
  \hbox to \hsize{
    \hss
    \includegraphics[width=0.6\hsize]{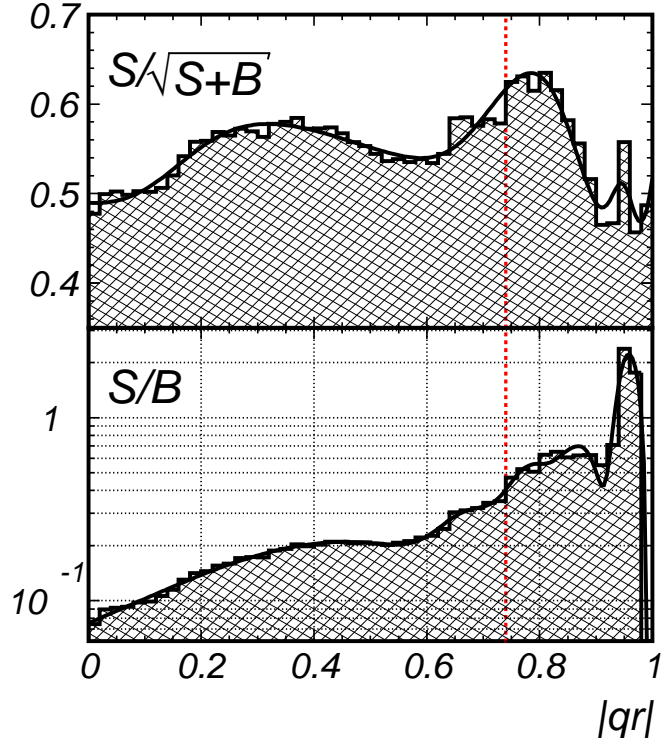}
    \hss
  }
  \caption{
    Effect of including $\left|qr\right|$ in the optimization procedure 
    described in the text.
    The variations in (top) $S/\sqrt{S+B}$ and (bottom) $S/B$
    with the $\left|qr\right|$ requirement are shown, after the 
    ${\mathcal L}_s/\left( {\mathcal L}_s + {\mathcal L}_b \right) > 0.9$
    selection.
    The line indicates the selection requirement that is obtained
    from the optimization procedure ($\left|qr\right| > 0.74$).
    Note that the statistical error on the entries becomes significant
    for values of $\left|qr\right|$ close to $1$.
    For this reason, a smoothed histogram curve is also shown to 
    guide the eye.
  }
  \label{fig:qr}
\end{figure}

We use a large statistics sample of $\rho^0\pi^0$ MC and data 
from a continuum dominated sideband region 
($5.23 < M_{\rm bc}/{\rm GeV}/c^2 < 5.26$ and 
$0.2 < \Delta E/{\rm GeV} < 0.4$),
in order to simultaneously find the optimum selection requirements 
on $\left|qr\right|$ and 
${\mathcal L}_s/\left( {\mathcal L}_s + {\mathcal L}_b \right)$, 
as defined above.
We find the sensitivity to $\rho^0\pi^0$ is maximized by 
requiring $\left|qr\right| > 0.74$ and 
${\mathcal L}_s/\left( {\mathcal L}_s + {\mathcal L}_b \right) > 0.9$,
where the branching fraction of 
$B^0 \to \rho^0\pi^0$ is taken to be $1 \times 10^{-6}$ 
as input to the optimization procedure.
The effect of the $\left|qr\right|$ requirement is shown in Fig.~\ref{fig:qr}.
Some modest improvement is seen in $S/\sqrt{S+B}$, 
whilst $S/B$ increases dramatically.

As for $\rho^{\pm}\pi^{\mp}$, 
we obtain the signal yield by fitting the $\Delta E$ 
distribution after requiring events be in the $M_{\rm bc}$ signal region
($M_{\rm bc} > 5.269~{\rm GeV}/c^2$),
and cross-check the result by fitting the $M_{\rm bc}$
distribution after requiring events be in the $\Delta E$ signal region
($-0.135 < \Delta E/{\rm GeV} < 0.080$).
We model the signal using a Crystal Ball lineshape,
with parameters determined from MC, and calibrated using control samples of
$B^- \to D^0 \rho^-$, with $D^0 \to K^-\pi^+$, $\rho^- \to \pi^-\pi^0$
and $\bar{B}^0 \to D^{*+}\rho^-$, with $D^{*+} \to D^0\pi^+$
and the same decays of the $D^0$ and $\rho$.  
In both cases the $\pi^0$ is required to have momentum in the CM frame
greater than $1.8~{\rm GeV}/c$, in order to have similar dynamics 
to the $\rho^0\pi^0$ final state.
We include additional components in the fit to accomodate 
continuum background (modelled by a Chebyshev polynomial
with parameters obtained from a fit to the sideband region),
$\rho^+\rho^0$ (shape from smoothed histogram of MC events)
and $\rho^+\pi^0$ (shape from smoothed histogram of MC events).
Contributions from $b \to c$ transitions are negligible 
in the fitted range of $\Delta E$, 
whilst the tiny contributions possible from other rare $B$ decays
are accounted for in the systematic error.
The only free parameters in the fit are again the signal and continuum
normalizations; the $\rho^+\rho^0$ yield and polarization are
fixed according to our recent measurements~\cite{jingzhi},
while the $\rho^+\pi^0$ yield is fixed based on theoretical expectations.

\begin{figure}
  \hbox to \hsize{
    \hss
    \includegraphics[width=0.45\hsize]{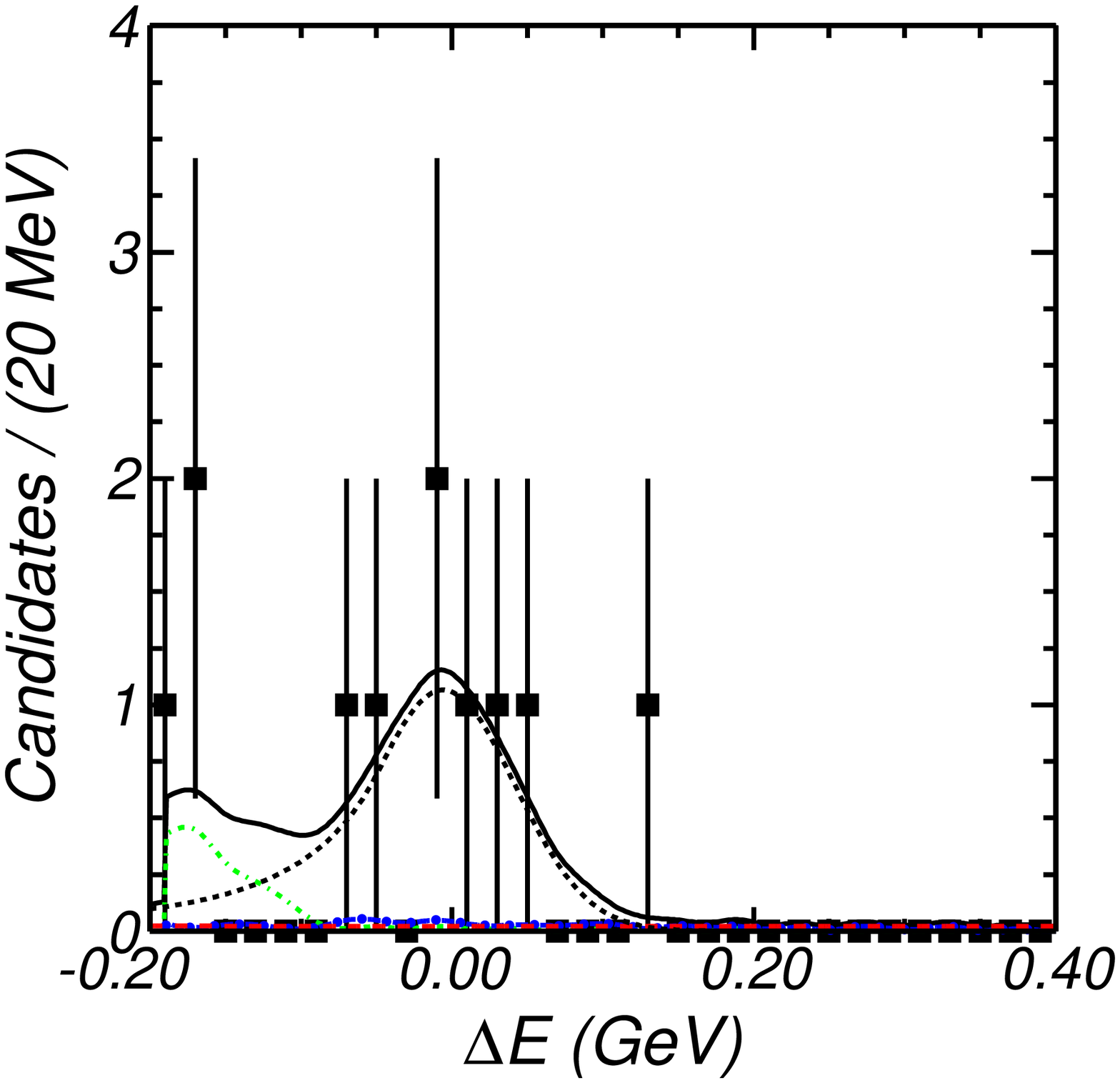}
    \hss
    \includegraphics[width=0.45\hsize]{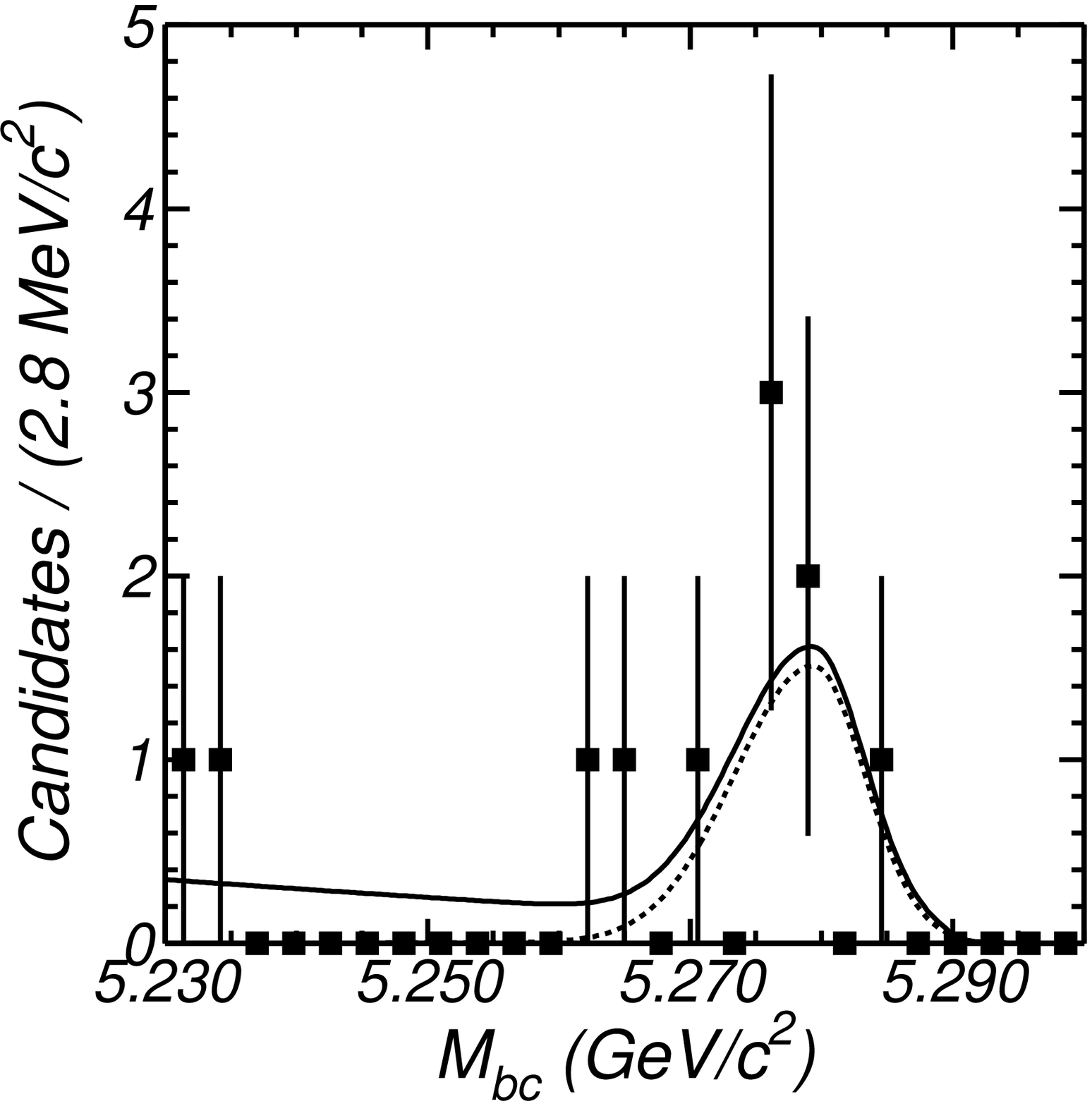}
    \hss
  }
  \caption{
    Fits to (left) $\Delta E$ and (right) $M_{\rm bc}$ 
    for $\rho^0\pi^0$ candidates.
    In the $\Delta E$ fit, contributions from signal (dashed line)
    and $\rho^+\rho^0$ (dot-dashed line) can be seen;
    the other contributions are very small.
    In the $M_{\rm bc}$ fit only contributions from signal (dashed line)
    and continuum are allowed.
    The sums of contributions are shown as solid lines.
  }
  \label{fig:rho0pi0_fit}
\end{figure}

The fit result is shown in Fig.~\ref{fig:rho0pi0_fit}.
The $\Delta E$ fit gives a signal yield of $6.6^{+3.2}_{-2.6}$
with a significance of $3.1 \sigma$.
The $M_{\rm bc}$ fit, in which only components for signal and continuum
are included,
gives a yield of $6.4^{+3.1}_{-2.4}$ with $3.6 \sigma$ significance.

\begin{figure}
  \hbox to \hsize{
    \hss
    \includegraphics[width=0.45\hsize]{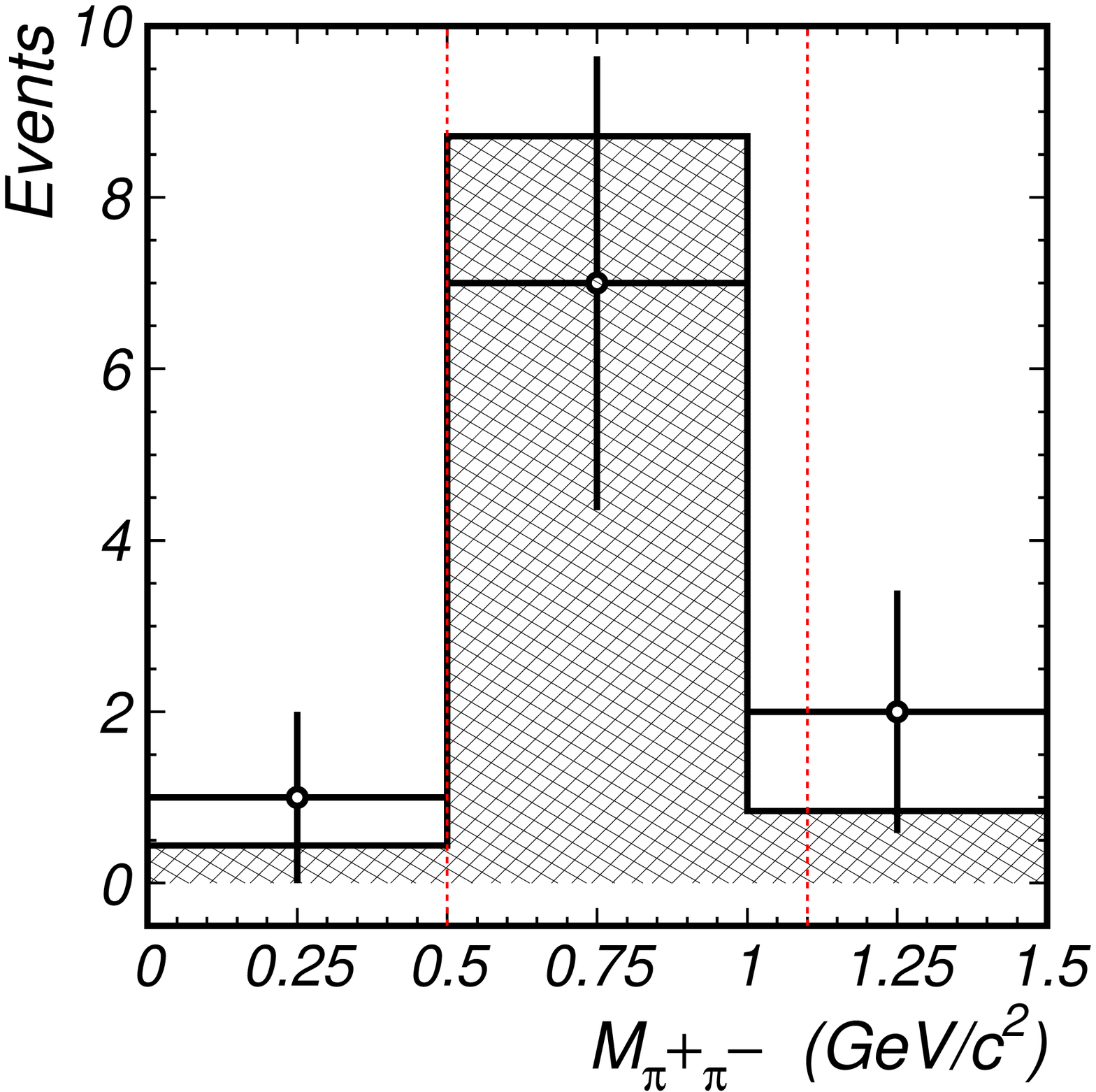}
    \hss
    \includegraphics[width=0.45\hsize]{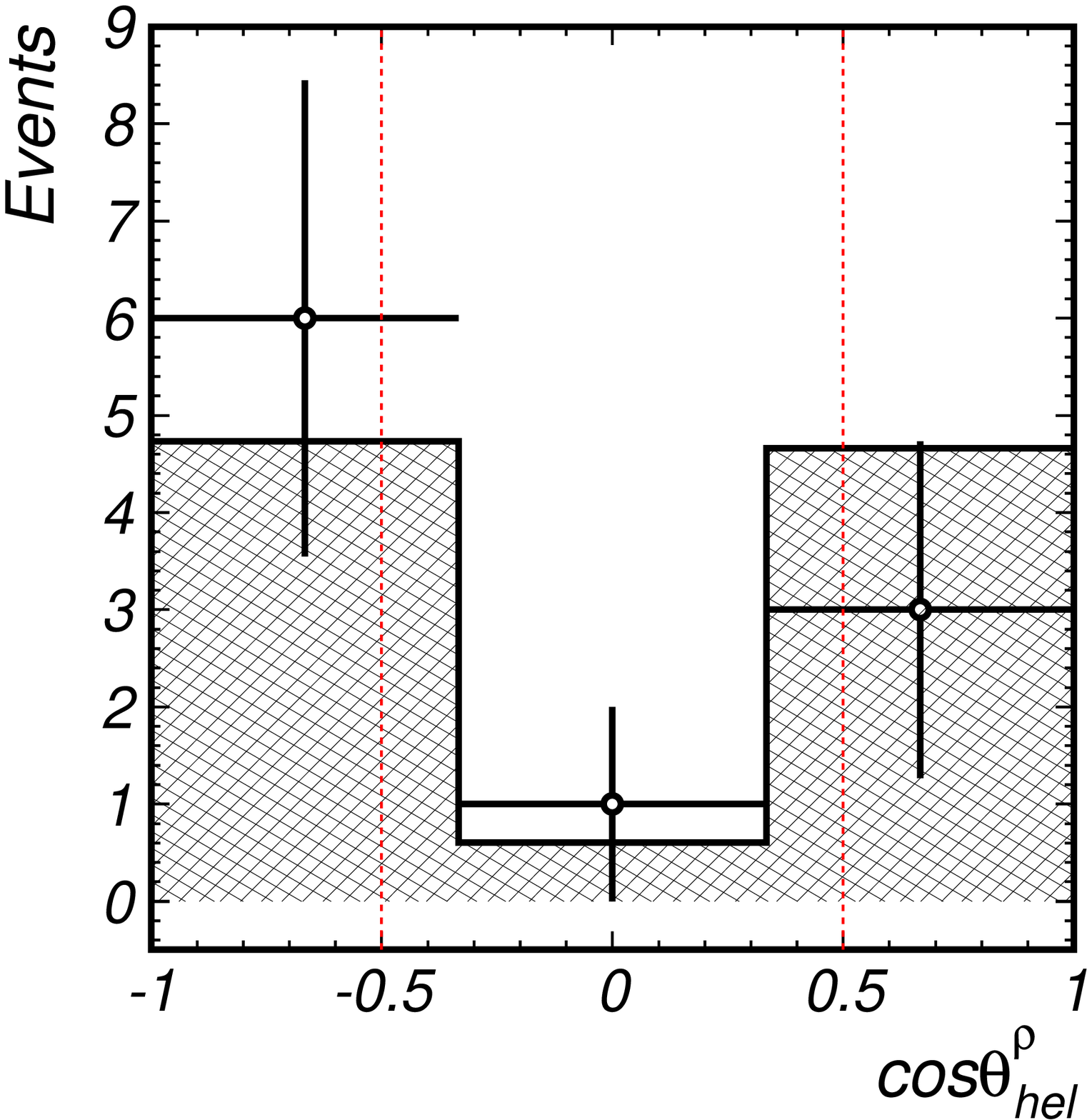}
    \hss
  }
  \caption{
    Distributions of (left) $M_{\pi^+\pi^-}$ and 
    (right) $\cos \theta^{\rho}_{\rm hel}$ for $\rho^0\pi^0$ candidate events.
    The selection requirements described in the text are shown 
    as dashed lines.
  }
  \label{fig:rho0pi0_check}
\end{figure}

In order to check that the signal candidates originate from
$B^0 \to \rho^0\pi^0$ decays, and not from either 
non-resonant $\pi^+\pi^-\pi^0$ or $\sigma \pi^0$,
we relax the criteria on $M_{\pi^+\pi^-}$ and $\cos \theta^{\rho}_{\rm hel}$
in turn and look at the distribution of the candidate events in
both $\Delta E$ and $M_{\rm bc}$ signal regions in those variables.
These distributions are shown in Fig.~\ref{fig:rho0pi0_check}.
Whilst the statistics are too small to make quantitative statements,
there is no evidence for any contribution other than $\rho^0\pi^0$.
We also consider possible contamination from $\rho^{\pm}\pi^{\mp}$;
from a large MC sample in which interference between resonances 
in not simulated, we expect $<0.1$ events to pass the selection requirements.

To obtain the branching fraction, we measure the efficiency using MC,
and correct for the pion identification efficiency as above.
The systematic error due to pion identification is $\pm 3\%$.
We also correct for possible differences between data and MC 
due to the $\left|qr\right|$ and 
${\mathcal L}_s/\left( {\mathcal L}_s + {\mathcal L}_b \right)$ selections;
the statistical errors of the control samples 
($\bar{B}^0 \to D^{*+}\rho^-$ with 
$D^{*+} \to D^0\pi^+$, $D^0 \to K^-\pi^+$, $\rho^- \to \pi^-\pi^0$ and
$B^- \to D^0\rho^-$ with $D^0 \to K^- \pi^+$, $\rho^- \to \pi^-\pi^0$
respectively) account for the largest contribution to the systematic error
($\pm 18\%$).
We also calculate systematic errors due to:
PDF shapes, by varying parameters by $\pm 1\sigma$ ($\pm 3\%$);
$\pi^0$ reconstruction efficiency, from the inclusive $\eta$ study ($\pm 4\%$);
track finding efficiency, from the partially reconstructed $D^*$ study 
($\pm 2\%$).
We repeat the fit after changing the normalization of the other $B$ decay
components according to the error in their branching fractions,
and obtain systematic errors from the change in the result.
For the unobserved mode $\rho^+\pi^0$ we
vary the normalization by a factor of two.
The total systematic eror due to rare $B$ decays is $\pm 5\%$,
we also verify that in the case that all rare $B$ contributions are 
simulaneously increased to their maximum values, 
the statistical significance remains above $3\sigma$.
The total systematic error is $\pm 20\%$, 
and we measure the branching fraction of $B^0 \to \rho^0\pi^0$ to be
\begin{equation}\nonumber
  {\mathcal B}\left( B^0 \to \rho^0\pi^0 \right) =
  \left( 6.0 ^{+2.9}_{-2.3} ({\rm stat}) \pm 1.2 ({\rm syst}) \right)
  \times 10^{-6}.
\end{equation}

In order to test the robustness of this result,
a number of cross-checks are performed.
We vary the selection requirements on $\left|qr\right|$ and 
${\mathcal L}_s/\left( {\mathcal L}_s + {\mathcal L}_b \right)$.
In all cases, consistent central values for the branching fraction 
are obtained.
We also select $\rho^{\pm}\pi^{\mp}$ candidates after using
this continuum rejection technique, and measure 
a branching fraction for $B^0 \to \rho^{\pm}\pi^{\mp}$ which is 
consistent with that reported above.
Furthermore, adopting the continuum rejection technique of
our $\rho^{\pm}\pi^{\mp}$ analysis and selecting $\rho^0\pi^0$ candidates
also results in a consistent central value for the branching fraction of
$B^0 \to \rho^0\pi^0$, although the signal is insignificant above the 
large continuum background.

\vspace{2ex}

In summary, we have measured the branching fraction 
\begin{equation}\nonumber
  {\mathcal B}\left( B^0 \to \rho^\pm \pi^\mp \right) = 
  \left( 29.1 ^{+5.0}_{-4.9}({\rm stat}) \pm 4.0({\rm syst}) \right) 
  \times 10^{-6},
\end{equation} 
in agreement with previous measurements.
We also measure the untagged asymmetry to be
\begin{equation}\nonumber
  {\mathcal A} =
  -0.38^{+0.19}_{-0.21} ({\rm stat}) ^{+0.04}_{-0.05} ({\rm syst}),
\end{equation}
consistent with zero with the current statistical precision.
In addition, we observe evidence,
with $3.1\sigma$ statistical significance, for $B^0 \to \rho^0\pi^0$ 
with a branching fraction of
\begin{equation}\nonumber
  {\mathcal B}\left( B^0 \to \rho^0\pi^0 \right) =
  \left( 6.0 ^{+2.9}_{-2.3} ({\rm stat}) \pm 1.2 ({\rm syst}) \right)
  \times 10^{-6}.
\end{equation}

This is the first evidence for $B^0 \to \rho^0\pi^0$,
with a branching fraction higher than most predictions~\cite{snyder_quinn}.
This may indicate that some contribution to the amplitude is larger
than expected, which may complicate the extraction of $\phi_2$
from time-dependent analysis of $\rho^{\pm}\pi^{\mp}$.

We wish to thank the KEKB accelerator group for the excellent
operation of the KEKB accelerator.
We acknowledge support from the Ministry of Education,
Culture, Sports, Science, and Technology of Japan
and the Japan Society for the Promotion of Science;
the Australian Research Council
and the Australian Department of Industry, Science and Resources;
the National Science Foundation of China under contract No.~10175071;
the Department of Science and Technology of India;
the BK21 program of the Ministry of Education of Korea
and the CHEP SRC program of the Korea Science and Engineering
Foundation;
the Polish State Committee for Scientific Research
under contract No.~2P03B 01324;
the Ministry of Science and Technology of the Russian Federation;
the Ministry of Education, Science and Sport of the Republic of
Slovenia;
the National Science Council and the Ministry of Education of Taiwan;
and the U.S.\ Department of Energy.


\begin{thebibliography}{25}
\bibitem{cc}
The inclusion of charge conjugate decays is implied 
thoughout, except where explicitly excluded.
\end{thebibliography}
\end{document}